\documentclass[superscriptaddress,twocolumn,showpacs,aps]{revtex4}
\usepackage{graphicx}
\usepackage{epsfig}
\usepackage{amsmath}

\makeatother
\begin{document}

\title{Thermal collapse of a granular gas under gravity}

\author{Dmitri Volfson}
\affiliation{Bioengineering Department, University of California,
San Diego, La Jolla} \affiliation{Institute for Nonlinear Science,
University of California, San Diego, La Jolla, CA 92093-0402}

\author{Baruch Meerson}
\affiliation{Racah Institute of Physics, Hebrew University of
Jerusalem, Jerusalem 91904, Israel}

\author{Lev S. Tsimring}
\affiliation{Institute for Nonlinear Science, University of
California, San Diego, La Jolla, CA 92093-0402}

\begin{abstract}
Free cooling of a gas of inelastically colliding hard spheres represents a
central paradigm of kinetic theory of granular gases. At zero gravity the
temperature of a freely cooling \textit{homogeneous} granular gas follows a
power law in time. How does gravity, which brings inhomogeneity, affect the
cooling?  We combine molecular dynamics simulations, a numerical solution of
hydrodynamic equations and an analytic theory to show that a granular gas
cooling under gravity undergoes thermal collapse: it cools down to zero
temperature and condenses on the bottom of the container in a finite time.

\end{abstract}
 \maketitle

\begin{figure}
\includegraphics[scale=0.5]{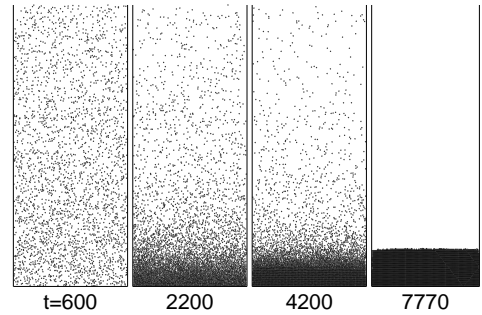}
\caption{Snapshots of an event-driven MD simulation at indicated
times for $N=5642$, $L_x=10^2$, $r=0.995$, $T_0=10$ and $g=0.01$.
Only a part of the box is shown. } \label{fig:MD}
\vskip-10pt
\end{figure}

\begin{figure}
\begin{tabular}{cc}
\includegraphics[scale=0.35]{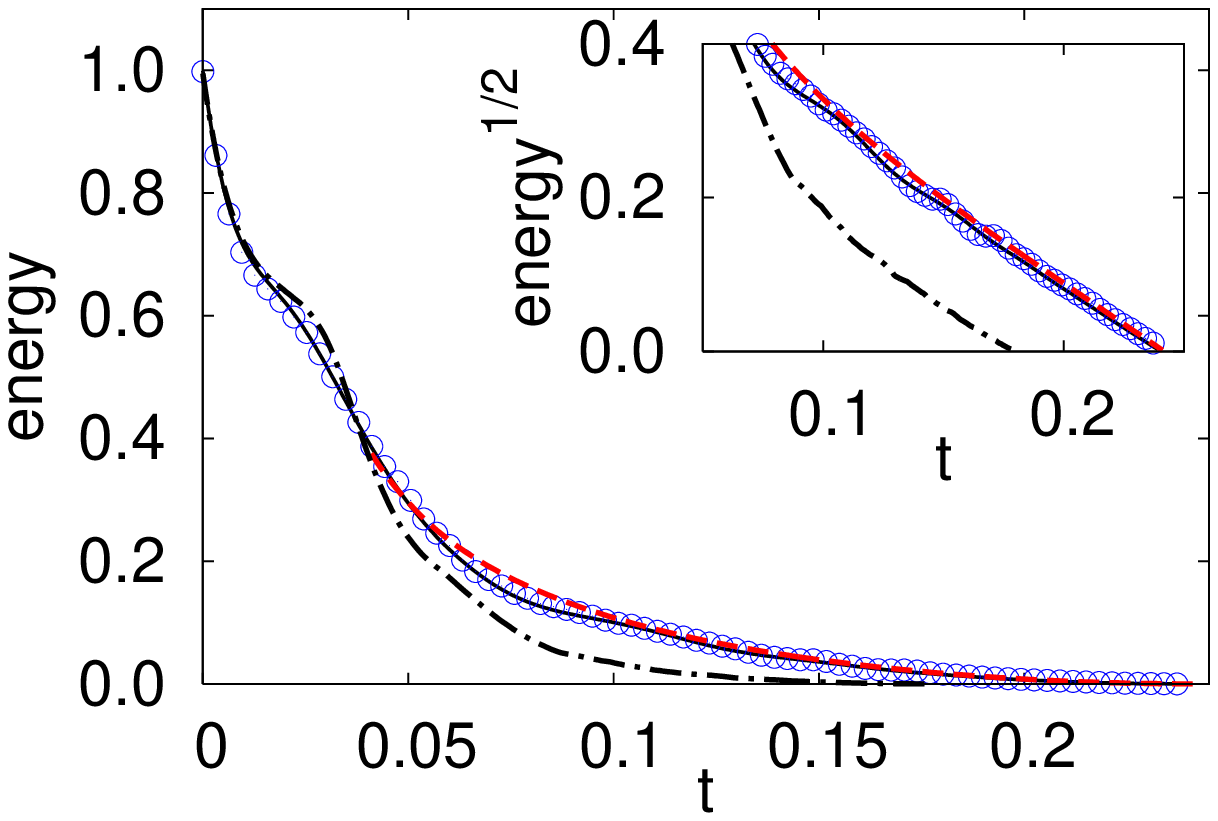}&
\includegraphics[scale=0.20]{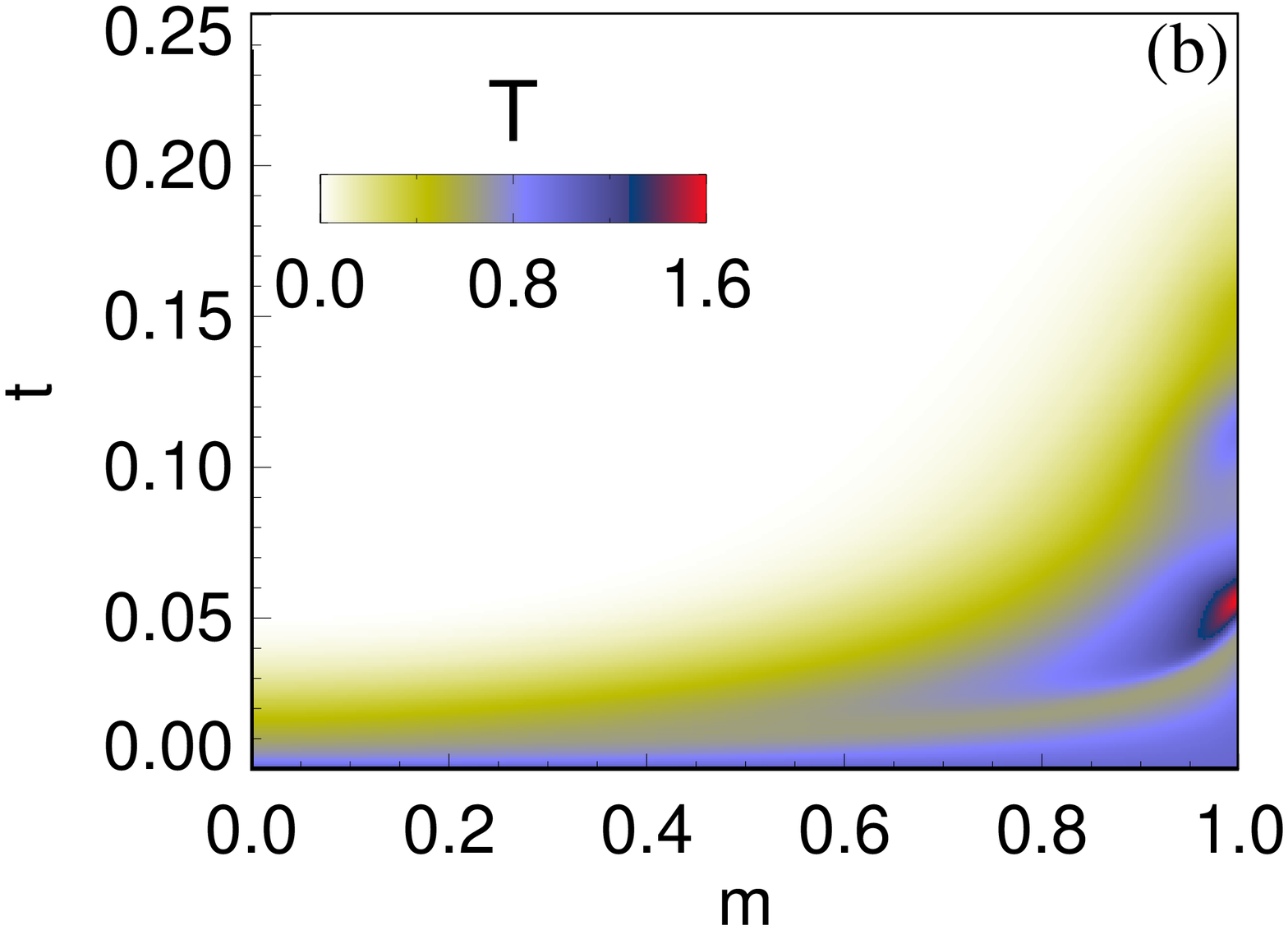}
\end{tabular}
\caption{(a) Total kinetic energy of the grains, normalized to its value at
$t=0$, vs. time for the run in Fig. \ref{fig:MD} (circles). Also shown are the
results from the full hydrodynamic model (\ref{nn})-(\ref{TT}) (black solid
line), from the $\omega$-equation (\ref{omega}) (red dashed line), and from a MD
simulation with a \textit{different}, non-isothermal initial condition (black
dash-dotted line). The respective hydrodynamic parameters are $\varepsilon =
10^{-2}$ and $\Lambda=5$. The inset shows the square root of the energy close to
thermal collapse. (b) a space-time plot of $T$ from the full hydrodynamic model
for the same simulation.} \label{fig:MD1} \vskip-10pt
\end{figure}

Granular gas, a low-density fluid of inelastic hard spheres, is a simple model
of granular flow, and it has attracted much attention from physicists
\cite{Haff,Goldhirsch+BP}. An undriven granular gas loses its kinetic energy via
inelastic collisions. In the Homogeneous Cooling State (HCS) the temperature $T$
of a dilute granular gas decays according to Haff's law \cite{Haff},
$T(t)=T_0(1+t/t_0)^{-2}$, where in two dimensions $t_0=\sqrt{\pi/2} \,(1-r^2)\,
d n_0 T_0^{1/2}$ is the cooling time, $n_0$ is the (constant) number density of
the particles, $d$ is the particle diameter and $r$ is
the coefficient of normal restitution. 
The HCS, and deviations from it, provide a rich testing ground for the ideas and
methods of kinetic theory of granular gases, and it has been investigated in
many theoretical works, see Ref. \cite{Goldhirsch+BP} and references therein.
Direct experimental observation on the HCS is difficult, not the least because
of gravity. Therefore it is somewhat surprising that there have been no
theoretical studies of the effect of gravity on the free cooling of
a granular gas. 
It is intuitively clear that gravity forces grains to sink to the bottom of the
container, where increased density enhances the collision rate and causes
``freezing" of the granulate. However, no quantitative analysis of this process
has ever been performed. Here we combine molecular dynamics (MD) simulations, a
numerical solution of granular hydrodynamic equations and analytical theory to
develop a detailed quantitative understanding of this cooling process. Our main
result is that, in a striking contrast to Haff's law, the gas undergoes thermal
collapse: it cools down to zero temperature and
condenses on the bottom plate in a finite time exhibiting, close to collapse, a
previously unknown universal scaling behavior. \\
\indent\textsf{MD simulations.} We employed an event-driven algorithm
\cite{Rapaport} to simulate a free cooling of an initially dilute gas of $N \gg
1$ identical nearly elastic, $1-r \ll 1$, hard disks of unit diameter and mass
in a two-dimensional container of width $L_x$ and infinite height. The (elastic)
bottom of the container is at $y=0$, the (elastic) side walls are at $x=0$ and
$L_x$. $L_x$ is chosen small enough so that any macroscopic structure in the
lateral direction is suppressed. The gravity acceleration $g$ acts in the
negative $y$
direction. 
Figure \ref{fig:MD} shows four snapshots of a typical simulation where, at
$t=0$, particles have a Maxwell velocity distribution, and a Boltzmann density
profile at constant temperature $T_0$ \cite{isothermal}. Collapse of all
particles to the bottom is observed at time $t_c=7770$. The circles in Fig.
\ref{fig:MD1}a show the time history of the simulated total kinetic energy of
the gas, normalized to its value at $t=0$. One can see that the total energy
drops to zero in a finite time. We observed a similar behavior in a wide range
of parameters, and also for a different, \textit{non-isothermal} initial state,
prepared by replacing the elastic bottom plate by a ``thermal" bottom plate
\cite{Rapaport} and waiting until a steady state is reached. In the latter case
the initial transient is somewhat different, but the energy decay law close to
collapse remains the same, see Fig. \ref{fig:MD1}a.

\textsf{Hydrodynamic theory}. The observed energy decay dynamics are remarkably
captured by hydrodynamic equations for the number density $n(y,t)$, vertical
velocity $v(y,t)$ and granular temperature $T(y,t)$. These equations are
systematically derivable from the Boltzmann equation generalized to account for
inelastic collisions of hard disks \cite{Goldhirsch+BP}. We assume a dilute gas,
an assumption which becomes invalid close to collapse. Following Ref.
\cite{bromberg03}, we rescale the variables using the gravity length scale
$\lambda=T_0/g$ and the heat diffusion time $t_d=\varepsilon^{-1}
(\lambda/g)^{1/2}$. The scaled parameter $\varepsilon= \pi^{-1/2}(L_x/Nd) \ll 1$
is of the order of the inverse number of layers of grains which form after the
particles settle on the bottom. The smallness of $\varepsilon$ guarantees that
$t_d$ is much longer than the fast hydrodynamic time $t_f = (\lambda/g)^{1/2}$.
We measure $v$ in units of $v_0=\lambda/t_d$, $n$ in units of $n_0=N/\lambda
L_x$, and $T$ in units of $T_0$. Furthermore, we exploit the one-dimensionality
of the flow and go over 
to Lagrangian mass
coordinate $m=\int_0^y n(y^{\prime},t) dy^{\prime}$ which varies between $0$ at
the bottom and $1$ (the total rescaled mass of the gas) as $y\to \infty$. 
The resulting rescaled hydrodynamic equations are
\cite{bromberg03,Fourier}:
\begin{eqnarray}
&&\partial_t (1/n) =\partial_m v\,,
\label{nn}\\
&&\varepsilon^2 \partial_t v =-\partial_m(nT)-1+(\varepsilon^2/
2)\,\partial_m(nT^{1/2}\partial_m v)\,,
\label{vv}\\
&&\partial_t T + nT\partial_m v = (\varepsilon^2/
2)\,nT^{1/2}(\partial_m v)^2+ \nonumber \\
&& (4/3)\,\partial_m(n\partial_m T^{3/2}) - 4\Lambda^2nT^{3/2}\,,
\label{TT}
\end{eqnarray}
In addition to
$\varepsilon$,  Eqs. (\ref{nn})-(\ref{TT}) include  the parameter
$\Lambda^2=\frac{1-r^2}{4\varepsilon^2}$ which shows the relative
role of the inelastic energy loss and heat diffusion. At the
boundaries $y=0$ and $\infty$ we demand zero fluxes of mass,
momentum and energy \cite{bromberg03}, 
which yield $v=\partial_m T = 0$ at $m=0$ and $n\partial_m v = n
\partial_m T = 0$ at $m=1$.

We solved Eqs. (\ref{nn})-(\ref{TT}) numerically in a wide range of parameters,
using a variable mesh/variable time step solver \cite{solver}.  The blue solid
line in Fig. \ref{fig:MD1}a depicts the sum of the thermal energy and
macroscopic kinetic energy of the gas versus time for the same parameters and
initial condition as in the MD simulation indicated by the circles. The
agreement is excellent, and thermal collapse is clearly observed
\cite{condensate}. At the very early stage of the cooling [with duration of
${\mathcal O}(\tau_f)$] we observed shock waves which form at large heights,
cause a transient heating of the gas there, and escape to $m=1$ ($y=\infty$),
see Fig. \ref{fig:MD1}b.

\textsf{Quasi-static flow.} If $\varepsilon \ll \min (1,\Lambda^{-2})$ then,
after the brief transient, a \textit{quasi-static} flow sets in. Here the
$\varepsilon^2$-terms in Eqs. (\ref{vv}) and (\ref{TT}) can be neglected, and
Eq. (\ref{vv}) reduces to the hydrostatic condition $\partial_m(nT)+1=0$ which
yields $nT=1-m$. Substituting $n=(1-m)/T$ into Eq. (\ref{TT}) and using Eq.
(\ref{nn}), we obtain a closed nonlinear equation for a new variable
$\omega(m,t)=T^{1/2}(m,t)$:
\begin{eqnarray}
\omega\partial_t \omega= \partial_m\left[(1-m)\partial_m\,
\omega\right] -\Lambda^2(1-m)\omega\,, \label{omega}
\end{eqnarray}
We will call Eq. (\ref{omega}) the $\omega$-equation; it was derived, in another
context, in Ref. \cite{bromberg03}. 

\begin{figure}
\begin{tabular}{cc}
\includegraphics[scale=0.35]{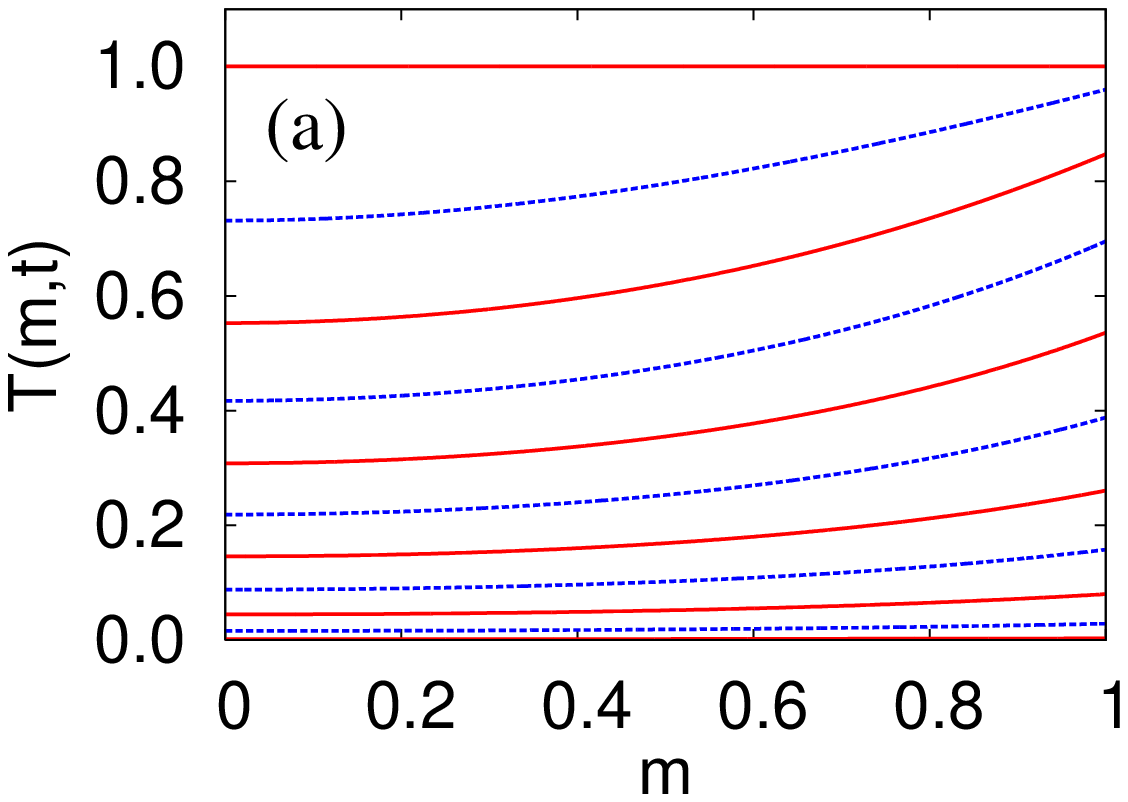}&
\includegraphics[scale=0.35]{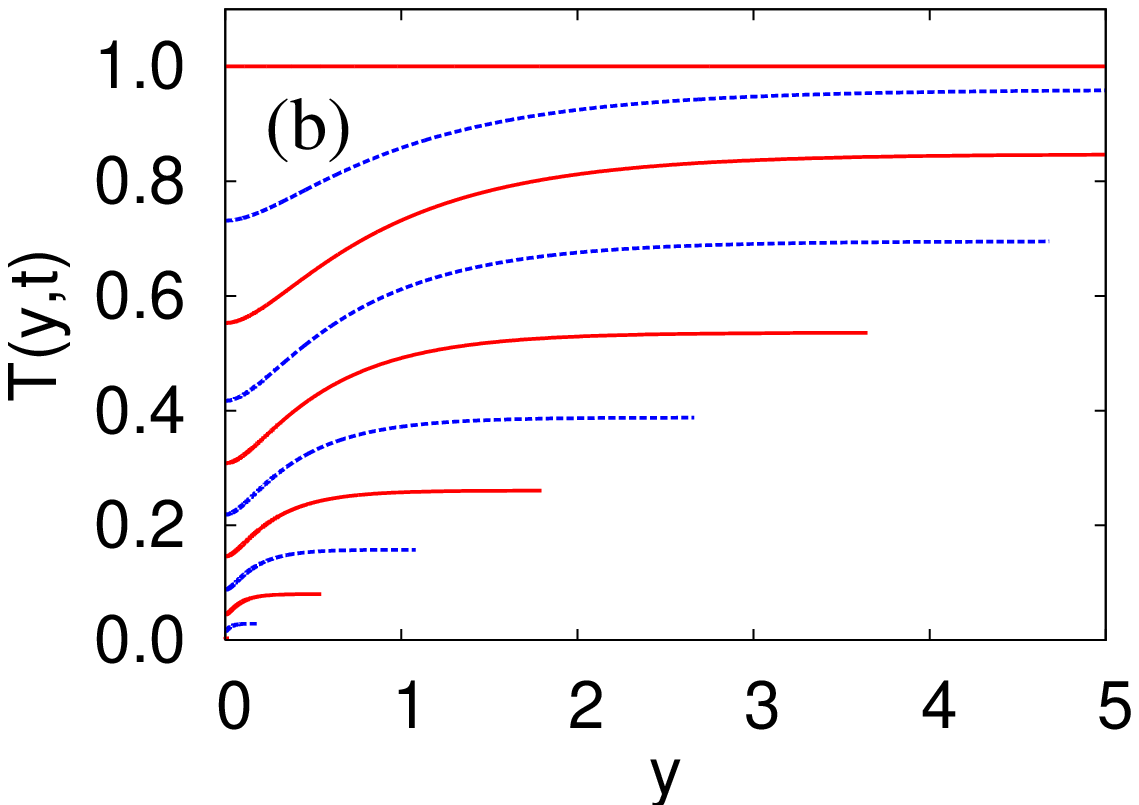}\\
\includegraphics[scale=0.35]{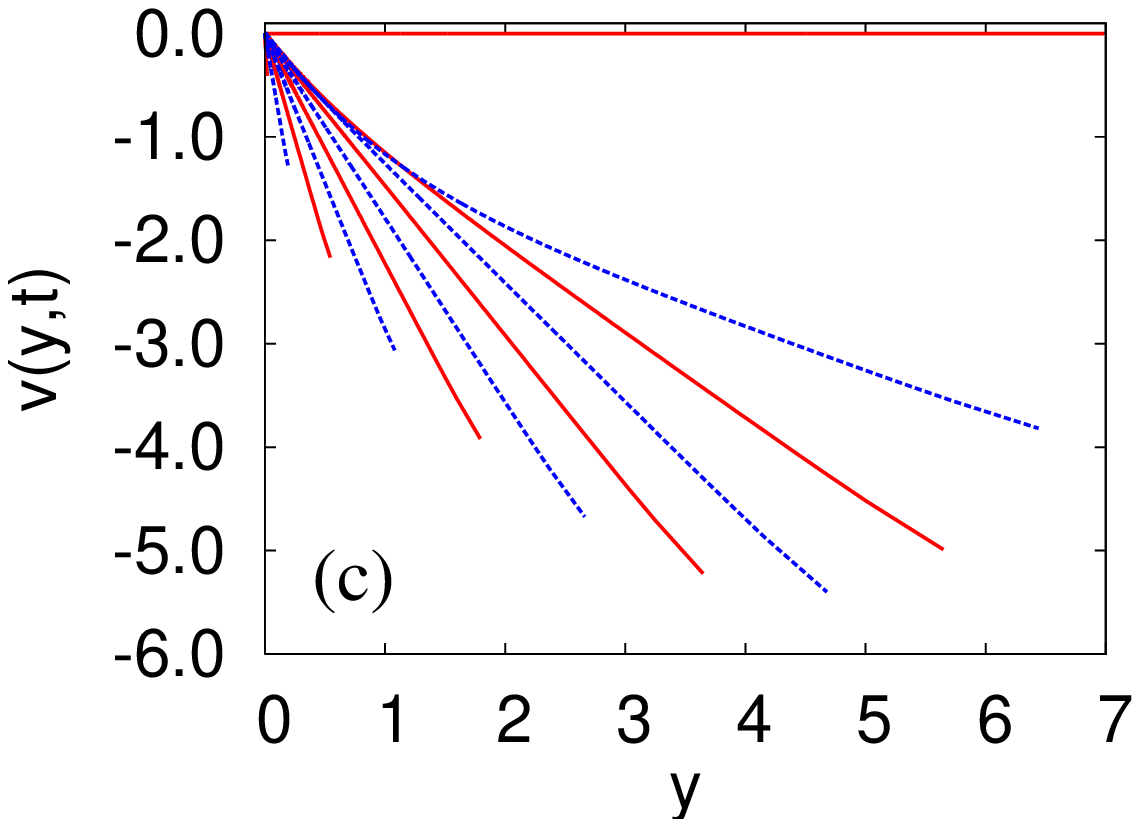}&
\includegraphics[scale=0.35]{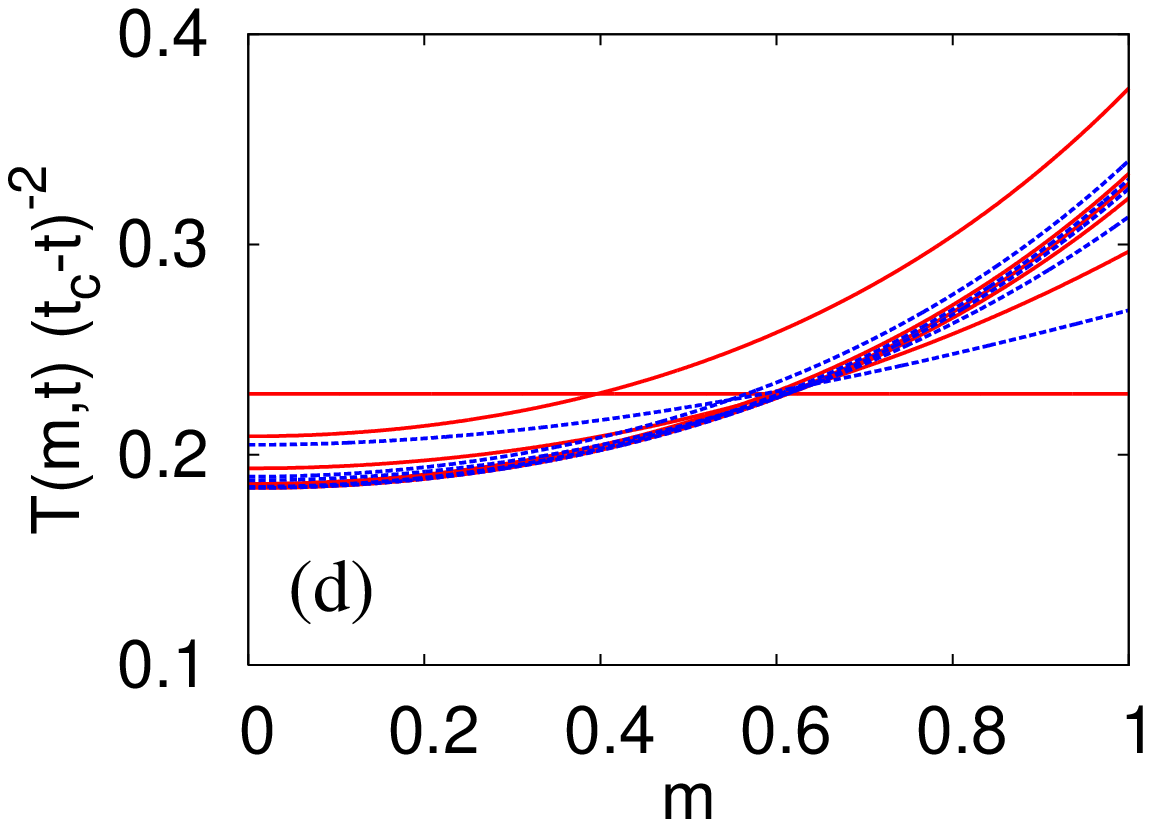}
\end{tabular}
\caption{Hydrodynamic fields predicted by the $\omega$-equation for
$\Lambda=1$: $T(m,t)$ at times separated by $\Delta t=0.2$ (a), and
$T(y,t)$ (b), $v(y,t)$ (c), and $T(m,t)(t_c-t)^{-2}$ (d) at the same
times. The late-time curves in d collapse into a single curve.}
\label{omega_m}
\vskip-10pt
\end{figure}

We solved the $\omega$-equation numerically [with the no-flux boundary
conditions $(1-m)\partial_m \omega=0$ at $m=0$ and $1$], using the same solver
\cite{solver}.
A typical example is shown in Fig. \ref{fig:MD1}a. Here we launched the
computation at scaled time $t=0.04$ when the hydrostatic condition $nT=1-m$
already holds well, and used the temperature profile, computed with the full
hydrodynamic solver, as the initial condition. One can see that the
$\omega$-equation provides a faithful description of the later stage of the
cooling. Figure \ref{omega_m} shows a different example of the cooling dynamics,
as described by the $\omega$-equation starting from $\omega(m,t=0)=1$. Here we
show the $T$- and $v$-profiles in both Lagrangian and Eulerian coordinates. In
all simulations thermal collapse is observed at a time $t_c$ which goes down as
$\Lambda$ increases. The collapse occurs simultaneously on the whole Lagrangian
interval $(0,1)$, see Fig. \ref{omega_m}a. As the density $n=(1-m)/T$ blows up
at $t=t_c$ at all $m\in [0,1)$, this Lagrangian interval corresponds to a single
Eulerian point $y=0$. Therefore, at time $t=t_c$
 \textit{all} of the gas condenses on the bottom plate $y=0$ and cools to a
zero temperature \cite{different}.

{\em Separable solution close to collapse.} As Fig. \ref{omega_m}d implies,
$\omega(m,t)$ becomes separable as $t \to t_c$. This remarkable solution can be
written as
\begin{equation}
\label{separable} \omega(m,t) = (t_c-t)\, Q(m)\,,
\end{equation}
where $Q(m)$ is determined by the nonlinear ODE
\begin{equation}
\left[(1-m)Q^{\prime}\right]^{\prime}-\Lambda^2(1-m)Q+ Q^2=0
\label{q}
\end{equation}
(the primes denote $m$-derivatives) and the boundary conditions
$(1-m)\,Q^{\prime} =0$ at $m=0$ and $1$. Function $Q(m)$ is uniquely
determined by $\Lambda$ and, at fixed $\Lambda$, can be found
numerically by shooting. In addition, we found $Q(m)$ perturbatively
for small and large $\Lambda$:

I. $\Lambda^2 \ll 1$. As it can be checked \textit{a posteriori}, in
this case $Q(m) ={\mathcal O} (\Lambda^2)$. Furthermore, as the heat
diffusion dominates over the inelastic energy loss, the solution
must be almost constant on the whole interval $0\le m<1$. Therefore,
we seek a solution in the form $Q(m)=\Lambda^2 Q_0 +\Lambda^4
Q_1(m)+ \Lambda^6 Q_2(m)+ \dots $.
Substituting this in Eq.
(\ref{q}) and equating terms of the same order in $\Lambda^2$, we
obtain the asymptotic solution
\begin{equation}
\label{qresult}
Q(m)=\frac{\Lambda^2}{2}-\frac{\Lambda^4}{16}+
\frac{\Lambda^4 m^2}{8}
+{\mathcal O}(\Lambda^6)\,.
\end{equation}
We checked that this solution is in excellent agreement with
numerical solutions of the $\omega$-equation at small $\Lambda$.

\begin{figure}
\begin{tabular}{cc}
\includegraphics[scale=0.36]{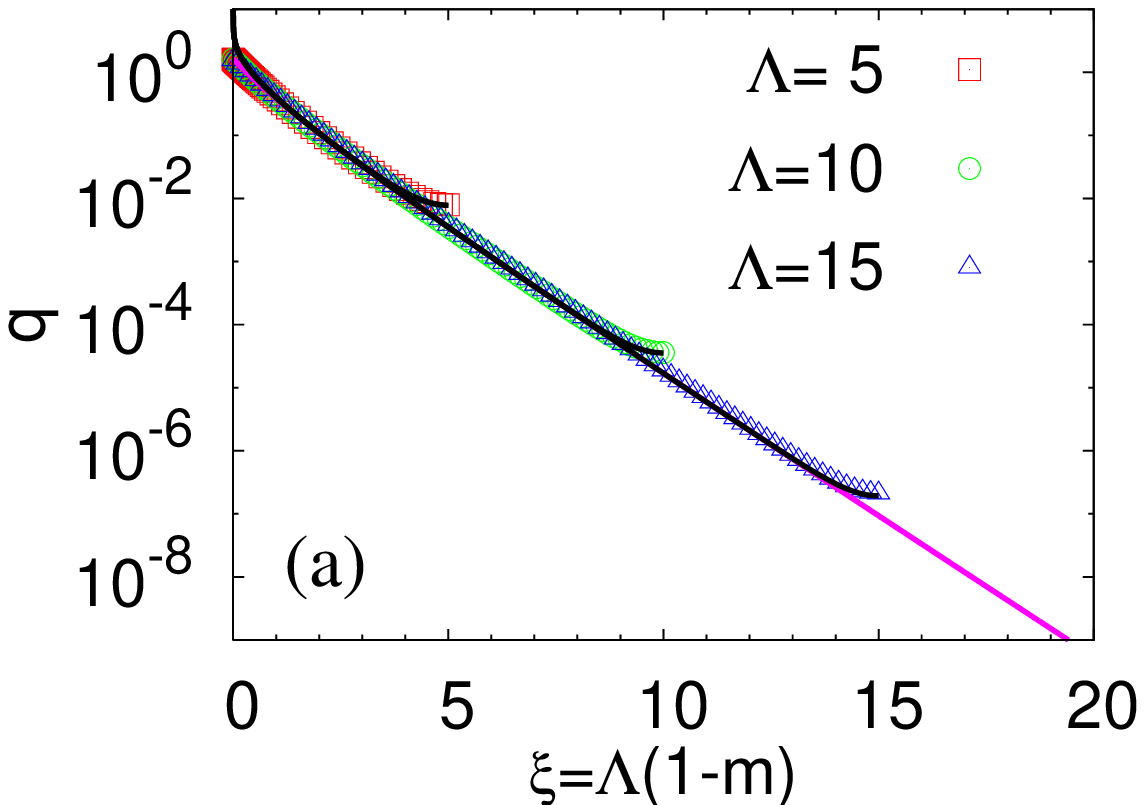}&
\hspace{-0.2in}\includegraphics[scale=0.36]{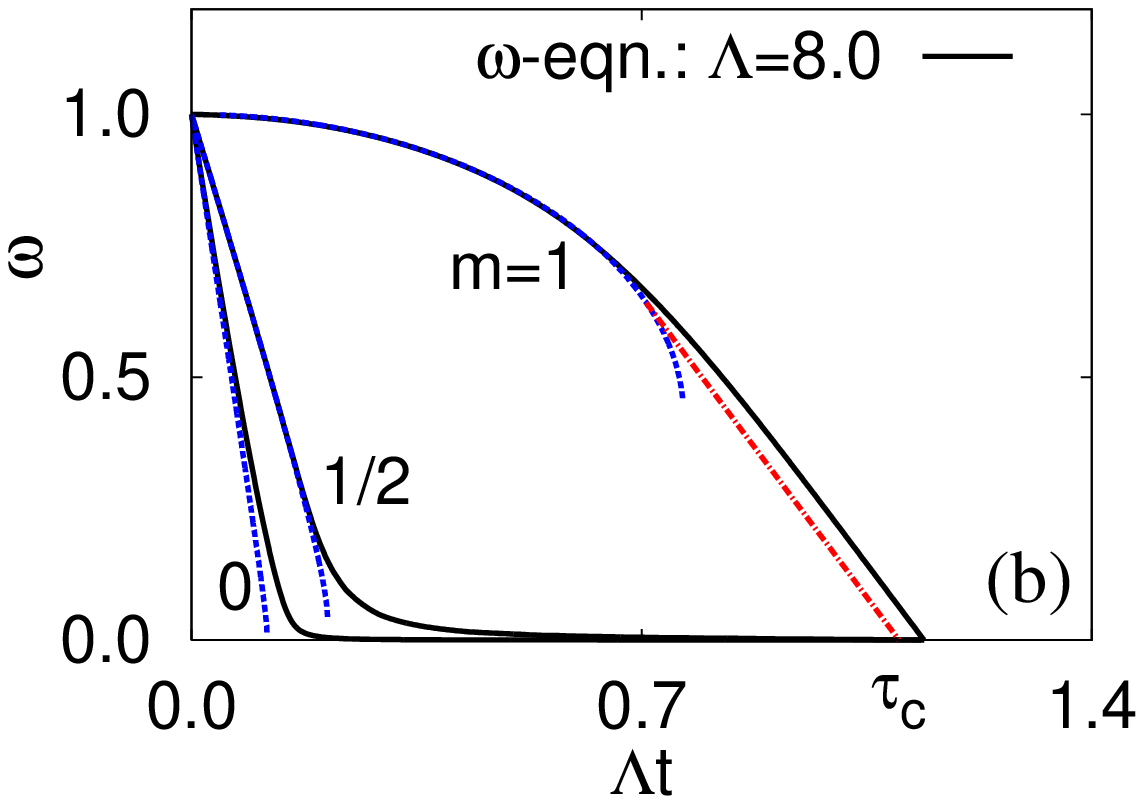}
\end{tabular}
\caption{(a) Numerical solutions of Eq. (\ref{qsc}) (symbols) and
the corresponding ``bulk solutions" $q_b(\xi)$ (lines) for
$\Lambda=5,10,15$. (b) The implicit solution (\ref{impl}) (the blue
lines) versus the numerical solution of the $\omega$-equation (the
black lines) for $\Lambda=8.0$ at three Lagrangian points: $m=0,
0.5$ and $1.0$. The red straight line shows the smoothly matched
asymptotic separable solution at $m=1$ (b).}\label{fig:qsc}
\vskip-10pt
\end{figure}

II. $\Lambda^2 \gg 1$. Here it is convenient to stretch the
Lagrangian coordinate, $\xi=\Lambda(1-m)$, and time $\tau=\Lambda
t$, so that $\Lambda$ drops from the $\omega$-equation
\begin{equation}
\label{omegasc} \omega\partial_{\tau} \omega = \xi
\partial_\xi^2 \omega+\partial_\xi\omega
-\xi\omega,
\end{equation}
but enters the integration interval $(0,\Lambda]$, whereas the
boundary condition are $\xi\partial_{\xi}\omega=0$. The separable
solution is $\omega(\xi,\tau)=(\tau_c-\tau) q(\xi)$, while the
boundary-value problem for $q(\xi)$ is
\begin{equation}
(\xi q^{\prime})^{\prime} -\xi q + q^2=0,\quad \xi
q^{\prime}=0\quad\mbox{at}\quad\xi=0,\Lambda\,. \label{qsc}
\end{equation}
At $\xi \gg 1$ $q(\xi)$ is exponentially small, so one can drop the
$q^2$-term and obtain $q_b(\xi) = C\,[K_0
(\xi)+K_1(\Lambda)I_1^{-1}(\Lambda)I_0(\xi)]$ (where $K_{0,1}(\xi)$
and $I_{0,1}(\xi)$ are the modified Bessel functions), which obeys
the boundary condition at $\xi=\Lambda$. This solution with
$C=0.951$ agrees well with the full numerical solution already at
$\xi>1$, see Fig. \ref{fig:qsc}, and therefore is valid everywhere
except the thin boundary layer at $m \to 1$. As the $q^2$-term
originates from the $\omega
\partial_{\tau}\omega$ term in Eq. (\ref{omegasc}), we realize that, almost
everywhere, the energy loss at late times is balanced by the heat
conduction, while the boundary layer at $m \to 1$ serves as a
dynamic ``bottleneck" of the cooling.

Outside of a thin boundary layer near $\xi=\Lambda$ (or $m=0$), the
solution is close to the solution of the same equation but on the
semi-infinite interval $(0,\infty)$. The latter one,
$q_\infty(\xi)$, is parameter-free and can be found numerically. The
shooting starts at the left boundary $\xi=0$ which is a regular
singular point of Eq. (\ref{qsc}). We demand that
$q_\infty^{\prime\prime}(\xi=0)$ be finite, which yields
$q_\infty^{\prime}(0)=-q_\infty(0)^2$. The shooting procedure gives
a unique value of $q_\infty(0)$ for which the solution does not
diverge toward $+\infty$ or $-\infty$ at large $\xi$. We find
$q_0\equiv q_\infty(0)=1.633356\dots$; the respective asymptotic
profile $q_\infty(\xi)$ is the envelope of the numerical profiles
$q(\xi)$ for different $\Lambda$ in Fig. \ref{fig:qsc}a.

{\it Early dynamics and collapse time.} To find the collapse time
$t_c$ [a free parameter in the separable solution
(\ref{separable})], one needs to solve the $\omega$-equation with a
given initial condition. In general, this can only be done
numerically. We obtained analytic estimates of $t_c$, separately for
small and large $\Lambda$, for $\omega(m,t=0)=1$.

For $\Lambda\ll1 $ the separable solution (\ref{separable}) and
(\ref{qresult}) is valid,  at order $\Lambda^2$, at any $t\ge 0$.
This yields the leading-order estimate $t_c \simeq 2\Lambda^{-2}$.
For $\Lambda \gg 1$ the initial stage of the cooling dynamics should
be addressed separately. We notice that at early times the term
$\xi\partial^2_\xi \omega$ in Eq. (\ref{omegasc}) is
small compared to the rest of terms.  
With this term neglected Eq. (\ref{omegasc}) reduces to a
first-order equation, $\omega\partial_\tau
\omega-\partial_\xi{\omega}= -\xi\omega$, 
which is soluble by characteristics. The solution
$\omega(\xi,\tau)$, in an implicit form, is
\begin{equation}\label{impl}
 \sqrt{2}\, e^{-\xi^2/2} \omega(\xi,\tau)
   \int_{\xi/\sqrt{2}}^{\sqrt{\frac{\xi^2}{2}-\ln
[\omega(\xi,\tau)]}} e^{z^2}\,dz = \tau\,;
\end{equation}
it is depicted, at points $m=0,1/2$ and $1$, in Fig.
\ref{fig:qsc}b. At $\xi \gg 1$ (where most of the gas is
located), Eq. (\ref{impl}) predicts an early-time asymptote
$\omega(\xi,\tau\ll1)=1-\xi \tau$ which can be also obtained
directly from Eq. (\ref{omegasc}) with the heat conduction neglected
completely. The ``bottleneck" of cooling, however, is at large
heights, $\xi \ll 1$, where the gas is very dilute. An early-time
asymptote there, as predicted from Eq. (\ref{impl}), is
\begin{equation}
\omega \simeq 1-\xi \tau - \tau^2/2 = 1-\Lambda^2 (1-m) t -\Lambda^2
t^2/2\,. \label{small_xi}
\end{equation}
The implicit solution (\ref{impl}) breaks down, at a given $\xi$, at
time $\tau\sim \mbox{min}\,(1,\xi^{-1})$, and then the full
$\omega$-equation must be solved. Eventually, as $\tau$ approaches
$\tau_c$, the separable solution (\ref{separable}) emerges.

\begin{figure} [ht]
\begin{tabular}{ll}
\includegraphics[scale=0.34]{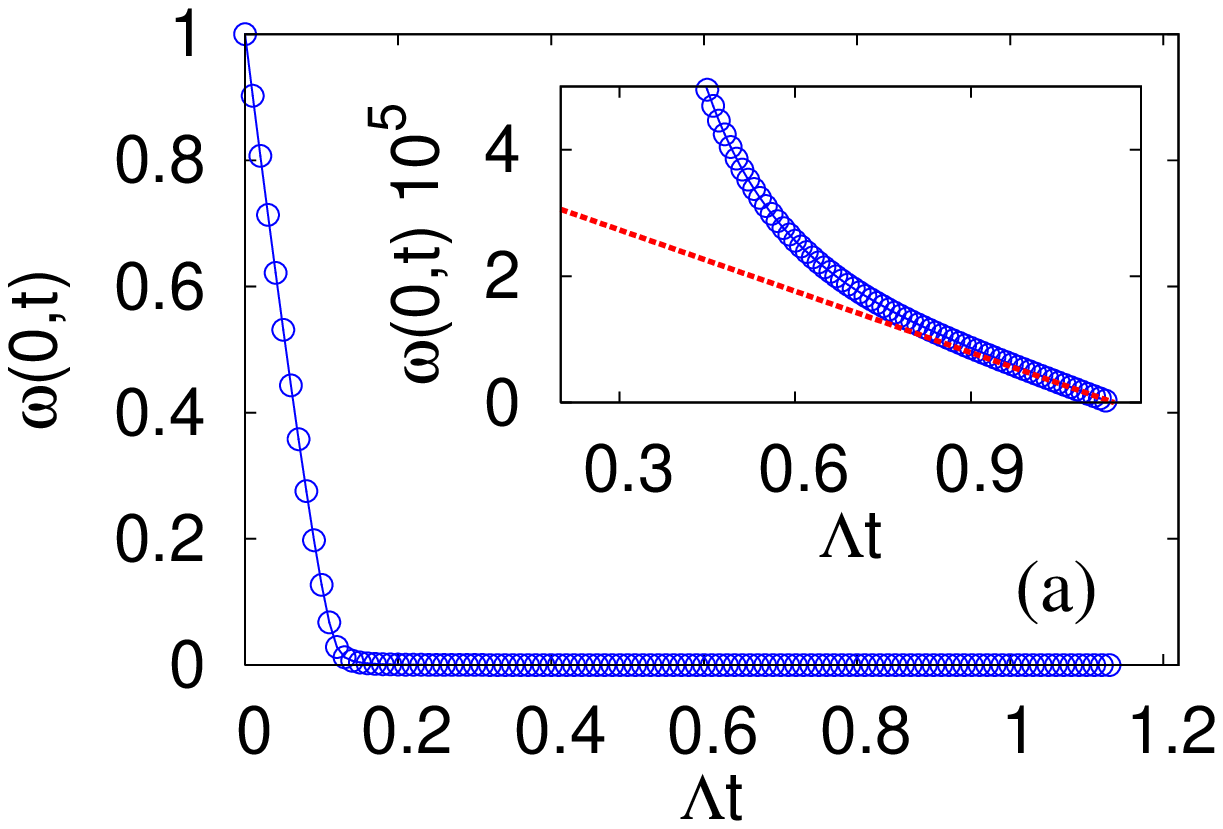}&
\includegraphics[scale=0.34]{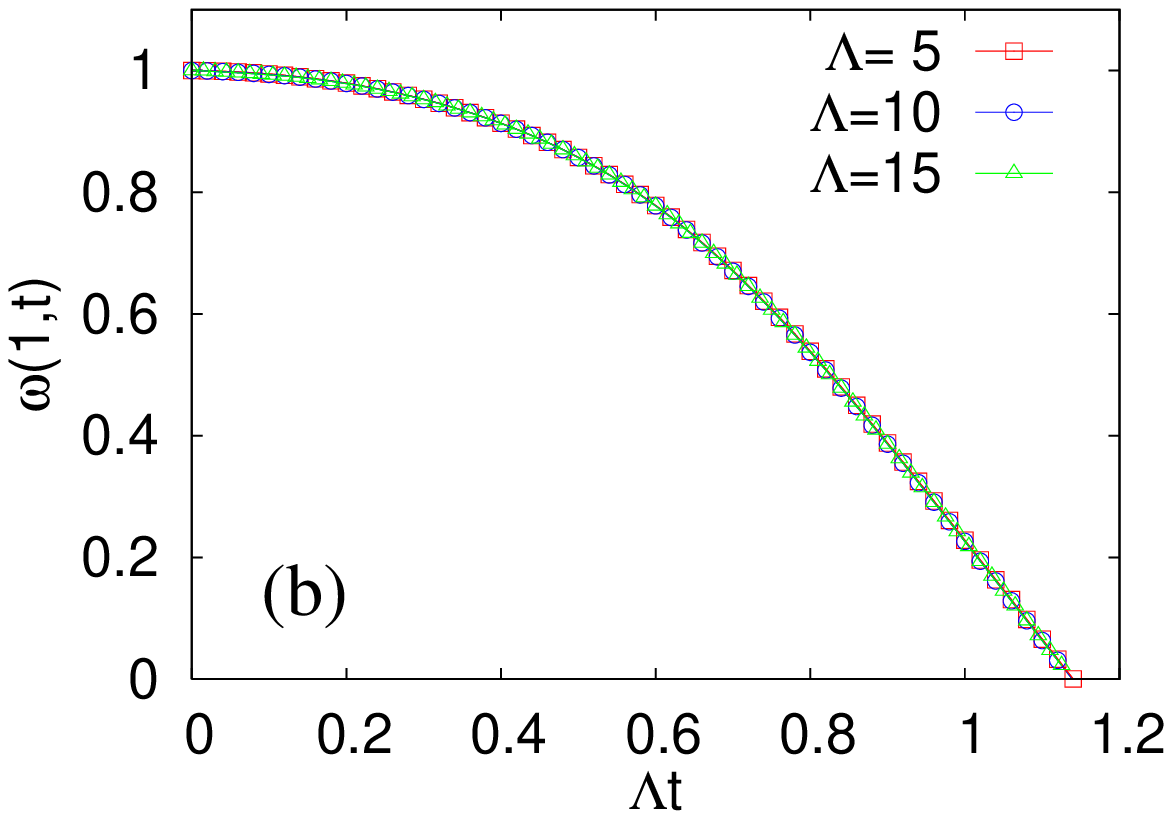}
\end{tabular}
\caption{(a) $\omega(m=0,t)$ at $\Lambda=10$. The inset shows a
blow-up near $\tau_c$, and the asymptote $(\tau_c-\tau) q(\Lambda)$
(the red dashed line). (b) $\omega(m=1,t)$ vs. $\tau=\Lambda t$ for
$\Lambda=5,10, 15$ collapse into a universal curve.}
\label{cooling2}
\vskip-20pt
\end{figure}

We stress that, at $\Lambda\gg 1$, the cooling process is highly
nonuniform, see Fig. \ref{cooling2}a and b. For example, at $m=0$ a
rapid initial decay $\omega=1-\Lambda^2 t$ crosses over, after a
short time $t \sim \Lambda^{-2}$,  into a very slow decay $\omega =
\Lambda q(\Lambda)(t_c-t)$, as $q(\Lambda)$ is exponentially small.
Meanwhile, at $m=1$ a slow initial decay $\omega=1-\tau^2/2$ crosses
over into a rapid decay $\omega=q_0 (\tau_c-\tau)$. At $t=t_c$
$\omega$ vanishes at \textit{all} $\xi$ (that is, at $y=0$). Note
that the dynamics at $m=1$ are independent of $\Lambda$ in the
stretched time $\tau=\Lambda t$ 

A good estimate of the collapse time $\tau_c$ at large $\Lambda$ can
be obtained by matching, at $\xi=0$, the late-time asymptote
$(\tau_c - \tau) q_0$ with the early-time solution
(\ref{impl}), 
see Fig. \ref{fig:qsc}b. This yields an algebraic equation for
$\tau_c$:
\begin{equation}
\left[(\pi/2)^{1/2}{\rm erfi}(2^{-1/2}\tau_c^{-1})+q_0^{-1}\right]=
\tau_c \exp[1/(2\tau_c^2)] \label{taus}
\end{equation}
which has a unique solution $\tau_c\simeq1.10$, or $t_c\simeq
1.10\Lambda^{-1}$. For comparison, a numerical solution of Eq.
(\ref{omegasc}), for large $\Lambda$, with the initial condition
$\omega(\xi,t=0)=1$  yields $\tau_c = 1.145$ which agrees with our
estimate to about $4\%$. The slope of the collapsed curves
$\omega(\xi=0,\tau)$ for large $\Lambda$ (Fig. \ref{fig:qsc}b) near
$\tau=\tau_c$, $\partial_\tau\omega(\xi=0,\tau_c)=1.624$, is in very
good agreement with the asymptotic value $q_0=1.633356$.
\begin{figure} [ht]
\includegraphics[scale=0.4]{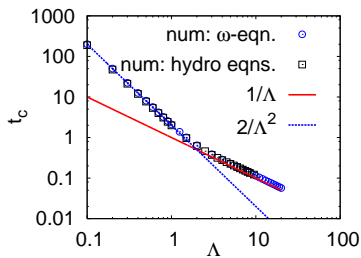}
\caption{The collapse time versus $\Lambda$. Numerical results from
the $\omega$-equation and from the full hydrodynamic equations
(\ref{nn})-(\ref{TT}) with $\varepsilon=0.025$ are shown by symbols,
the small- and large-$\Lambda$ asymptotes are denoted by
lines.} \label{fig:Tcol_Lmb}
\vskip-10pt
\end{figure}
Our predictions of the $\Lambda$-dependence of  $t_c$ are summarized
in Fig. \ref{fig:Tcol_Lmb}. The small- and large-$\Lambda$
asymptotes are in excellent agreement with numerical results.
Returning to the dimensional units, we observe that, at $\Lambda \ll
1$ the collapse time $t_c$ is much longer than the heat diffusion
time. At $\Lambda \gg 1$ $t_c$ is of the order of $(\varepsilon
\Lambda)^{-1} t_f \sim (1-r^2)^{-1/2} t_f$ which, for nearly elastic
collisions, is much longer than the free fall time $t_f$.

Having found $\omega(\xi,\tau)$ we can find the rest of hydrodynamic
fields.  Here we present the results for $\Lambda \gg 1$. In the
early stage of cooling  the gas density is
$n(\xi,\tau)=(\xi/\Lambda)(1-\xi\tau)^{-2}$ and velocity
$v(\xi,\tau)=\Lambda(\Lambda-\xi)[(\Lambda+\xi) \tau-2]$. Close to
collapse the density
$n(\xi,\tau)=\xi(\tau_c-\tau)^{-2}\,q^{-2}(\xi)$ blows up as
$(\tau_c-\tau)^{-2}$. The gas velocity is $v(\xi,\tau)= - 2 \Lambda
(\tau_c-\tau) \int_{\xi}^{\Lambda}
[q^2(\xi^{\prime})/\xi^{\prime}]\,d\xi^{\prime}$. At $\tau<\tau_c$
it diverges logarithmically at $\xi=0$ (that is, linearly at $y \to
\infty$), but vanishes everywhere at $\tau=\tau_c$, while the mass
flux $nv$ blows up at $\tau=\tau_c$. Going back to Eulerian
coordinate $y=(\tau_c-\tau)^2 \int_{\xi}^{\Lambda}
[q^2(\xi^{\prime})/\xi^{\prime}]\,d\xi^{\prime}$, we see that the
velocity field is simply $v(y,t)=-2 y/(t_c-t)$.

\textsf{Summary.} Our MD simulations and hydrodynamic theory depict a coherent
picture of thermal collapse which develops in the process of a free cooling of a
granular gas under gravity. One of the signatures of this picture is the
universal scaling
behavior of the total energy $E(t) \sim (t_c-t)^2$ as $t \to t_c$. 

It would be interesting to test the quantitative predictions of our theory in
experiment. A possible experiment can employ metallic spheres rolling  on a
slightly inclined smooth surface and driven by a rapidly vibrating bottom wall,
like in Ref. \cite{Kudrolli}. After the ``granular gas" reaches a steady state,
one stops the driving and follows the cooling dynamics with a fast camera and a
particle tracking software. While particle rotation and rolling friction may
prove important, we expect that the main predictions of the theory, including
the scaling behavior of the total energy at $t \to t_c$, will persist.

LT and DV are grateful to the U.S. Department of Energy for financial support
(Grant DE-FG02- 04ER46135). BM acknowledges support from the Israel Science
Foundation (grant No. 107/05) and from the German-Israel Foundation for
Scientific Research and Development (Grant I-795-166.10/2003). 


\begin{thebibliography}{9}
\bibitem{isothermal} In experiment, an (almost) isothermal initial state
is easily realized when the particle density $N/L_x$ and the inelasticity $1-r$
are not
too large, see e.g. Ref. \cite{Kudrolli}.
\bibitem{hydro} Granular hydrodynamics
assumes scale separation, which often demands nearly elastic collisions
\cite{Goldhirsch+BP}.
\bibitem{condensate} As the gas approaches collapse, a close-packed ``condensate" forms, where the
dilute gas model breaks down.  Therefore, it might seem surprising that it
describes the kinetic energy history remarkably well, see Fig. \ref{fig:MD1}a,
until the time of collapse. The reason is that the kinetic energy of the
condensate particles is extremely small, and the main
contribution to the energy comes from the gas particles.
\bibitem{Fourier} In the expression for the heat flux we omitted a term
with the \textit{density} gradient \cite{Goldhirsch+BP}. It
can be neglected in the nearly elastic limit.
\end{thebibliography}
\end{document}